\documentclass{ws-procs9x6}

\begin{document}

\title{Sfermions and gauginos in a Lorentz-violating theory}
\author{Roland E. Allen and Seiichirou Yokoo}
\maketitle

\address{Physics Department, Texas A\&M University \\
College Station, Texas 77843 \\
E-mail: allen@tamu.edu}

\abstracts{In Lorentz-violating supergravity, sfermions have spin 1/2 and
other unusual properties. If the dark matter consists of such particles,
there is a natural explanation for the apparent absence of cusps and other
small scale structure: The Lorentz-violating dark matter is cold because of
the large particle mass, but still moves at nearly the speed of light.
Although the R-parity of a sfermion, gaugino, or gravitino is +1 in the
present theory, these particles have an ``S-parity'' which implies that
the LSP is stable and that they are produced in pairs. On the other
hand, they can be clearly distinguished from the superpartners of
standard supersymmetry by their highly unconventional properties.}

In Lorentz-violating supergravity\cite{allen-2002,allen-2003}, sfermions
have spin 1/2. For one $SO\left( 10\right) $ generation, there are $16$
fields which are initially massless and right-handed. If half of these are
converted to left-handed charge-conjugate fields\cite
{allen-yokoo-2003,allen-yokoo-2004}, one obtains Lagrangian densities of the
form
\begin{eqnarray}
\mathcal{L}_{R} &=&\frac{1}{2}\,\left( -\bar{m}^{-1}\eta ^{\mu \nu }\partial
_{\mu }\psi _{R}^{\dagger }\partial _{\nu }\psi _{R}+\psi _{R}^{\dagger
}i\sigma ^{\mu }\partial _{\mu }\psi _{R}\right) +h.c. \\
\mathcal{L}_{L} &=&\frac{1}{2}\,\left( -\bar{m}^{-1}\eta ^{\mu \nu }\partial
_{\mu }\psi _{L}^{\dagger }\partial _{\nu }\psi _{L}-\psi _{L}^{\dagger }i
\bar{\sigma}^{\mu }\partial _{\mu }\psi _{L}\right) +h.c.
\end{eqnarray}
where $\psi _{R}$ and $\psi _{L}$ are 2-component left-handed and
right-handed fields, with $\bar{\sigma}^{0}=\sigma ^{0}=\mathbf{1}$ and
$\bar{\sigma}^{k}=-\sigma ^{k}$ as usual, and $\eta ^{\mu \nu
}=diag(-1,1,1,1) $. (The spacetime coordinate system in which the initial
fields are right-handed is defined above (9) of Ref. 3.) One can change from
the dimension $3/2$ fields $\psi $ to conventional dimension $1$ bosonic
fields $\phi $ by absorbing a factor of $\bar{m}^{-1/2}$, but that would
have no effect in the following arguments. The energy scale $\bar{m}$ is not
determined by the present theory, but is assumed to lie well above $1$ TeV--
perhaps as high as $10^{9}$ TeV or even the Planck scale.

Because of the minus sign in (2), one cannot couple left-and right-handed
fields with a Dirac mass in the normal way. A Majorana mass is also not
appropriate (except possibly in the case of sneutrinos). In the present
paper we therefore consider a Lorentz-violating mass which is postulated to
arise from supersymmetry breaking, at some high energy scale, within the
present Lorentz-violating theory. The Lagrangians then become
\begin{eqnarray}
\mathcal{L}_{R}^{\prime } &=&\bar{m}^{-1}\eta ^{\mu \nu }\psi _{R}^{\dagger
}\partial _{\mu }\partial _{\nu }\psi _{R}+\psi _{R}^{\dagger }i\sigma ^{\mu
}\partial _{\mu }\psi _{R}-m_{R}\psi _{R}^{\dagger }\psi _{R} \\
\mathcal{L}_{L}^{\prime } &=&\bar{m}^{-1}\eta ^{\mu \nu }\psi _{L}^{\dagger
}\partial _{\mu }\partial _{\nu }\psi _{L}-\psi _{L}^{\dagger }i\bar{\sigma}
^{\mu }\partial _{\mu }\psi _{L}-m_{L}\psi _{L}^{\dagger }\psi _{L}
\end{eqnarray}
where the prime indicates that a mass has been added and integrations by
parts have been performed in the action $S=\int d^{4}x\,\mathcal{L}$. The
mass term is invariant under a rotation, but not under a boost, since the
transformation matrix for a boost is not unitary.

The equations of motion are
\begin{eqnarray}
\bar{m}^{-1}\eta ^{\mu \nu }\partial _{\mu }\partial _{\nu }\psi
_{R}+i\sigma ^{\mu }\partial _{\mu }\psi _{R}-m_{R}\psi _{R} &=&0 \\
\bar{m}^{-1}\eta ^{\mu \nu }\partial _{\mu }\partial _{\nu }\psi _{L}-i
\bar{\sigma}^{\mu }\partial _{\mu }\psi _{L}-m_{L}\psi _{L} &=&0.
\end{eqnarray}
As in Ref. 1, let each field be represented as
\begin{equation}
\psi =\sum_{n}\,a_{n}\psi _{n} \quad , \quad
\psi _{n}\left( \vec{x}\right) =A_{n}\;\chi _{n}\;e^{i\vec{p}\cdot \vec{x}}
\quad , \quad  a_{n}\left( t\right) =e^{-i\omega _{n}t}a_{n}\left( 0\right)
\end{equation}
where $n\leftrightarrow \vec{p},\lambda $ and there are two possibilities
for the 2-component spinor $\chi _{n}$:
\begin{equation}
\vec{p}\cdot \vec{\sigma}\,\chi _{R}\;=+\left| \vec{p}\right| \,\chi
_{R}\quad ,\quad \vec{p}\cdot \vec{\sigma}\,\chi _{L}\;=-\left| \vec{p}
\right| \,\chi _{L}.
\end{equation}
If we require only that each $a_{n}\psi _{n}$ satisfy the equation of motion
for $\psi $, there are $4$ solutions:
\begin{equation}
\pm 2\omega =-\bar{m}\pm \sqrt{\bar{m}^{2}+4p^{2}\pm 4\bar{m}
\left( p+m\right) }
\end{equation}
where $m$ represents either $m_{R}$ or $m_{L}$ and the three independent
$\pm $ signs have the following meaning: The first $\pm $ sign, on the left
side of the equation, is $+$ for a right-handed field $\psi _{R}$ and $-$
for a left-handed field $\psi _{L}$. The last $\pm $ sign, under the
radical, is $+$ for a right-handed solution containing $\chi _{R}$ and
$- $ for a left-handed solution containing $\,\chi _{L}$. Finally, the
middle $\pm $ sign, preceding the radical, indicates that there are two
solutions for a given $\psi $ and $\,\chi $, with the $+$ sign corresponding
to the normal solution, for which $\omega \rightarrow 0$ as $\left| \vec{p}
\right| \rightarrow 0$, and the $-$ sign to an extremely high energy
solution, for which $\left| \omega \right| \rightarrow \bar{m}$ as
$\left| \vec{p}\right| \rightarrow 0$.

We will now see that not all of the $4$ above solutions are physical,
because one of them may correspond to negative-norm states which are
inadmissible in a proper positive-norm Hilbert space. After discarding any
unphysical solution, however, we are still left with enough basis functions
$\psi _{n}\left( \vec{x}\right) $ to have a complete set of functions for (i)
representing an arbitrary classical field $\psi \left( 
\vec{x}\right) $ and
(ii) satisfying the quantization condition below. The canonical momenta are
(in the notation of Ref. 1)
\begin{eqnarray}
\pi _{\psi }^{\dagger } &=&\frac{\partial \mathcal{L}_{\psi }}{\partial
\dot{\psi}}=\;\bar{m}^{-1}\dot{\psi}^{\dagger }\pm \frac{1}{2}i\psi ^{\dagger }
=\frac{1}{2}i\sum_{n}\,\left( \pm 1+2\omega _{n}/\bar{m}\right)
a_{n}^{\dagger }\psi _{n}^{\dagger } \\
\pi _{\psi } &=&\frac{\partial \mathcal{L}_{\psi }}{\partial \dot{\psi}
^{\dagger }}=\bar{m}^{-1}\dot{\psi}\pm \frac{1}{2}i\psi =\frac{1}{2}
i\sum_{n}\,\left( \pm 1-2\omega _{n}/\bar{m}\right) a_{n}\psi _{n}
\end{eqnarray}
where the upper sign holds for a right-handed field and the lower for a
left-handed field. We again quantize by interpreting $\psi $ and $\pi
^{\dagger }$ as operators, and requiring that
\begin{equation}
\left[ \psi _{\alpha }\left( \vec{x},t\right) ,\pi _{\psi \beta }^{\dagger
}\left( \vec{x}\,^{\prime },t\right) \right] _{-}=i\delta \left( \vec{x}-
\vec{x}\,^{\prime }\right) \delta _{\alpha \beta }
\end{equation}
where $\alpha $ and $\beta $ label the two components of $\psi $ and $\pi
_{\psi }^{\dagger }$. Following the same logic as in Ref. 1, one
finds that this requirement can be satisfied if
\begin{equation}
2\bar{m}^{-1}\pm \omega _{n}^{-1}>0 .
\end{equation}
Depending on the 3-momentum $\vec{p}$, there
are either three or four solutions which meet this condition (and which
are therefore included in the representation of the physical field operator
$\psi _{R}$ or $\psi _{L}$):
\bigskip

\noindent
\begin{tabular}{|l|l|l|}
\hline
$R$ & $\pm 2\omega =-\bar{m}+\sqrt{\bar{m}^{2}+4\left| \vec{p}\right| ^{2}
+4\bar{m}\left( \left| \vec{p}\right| +m\right) }$ & all $\left| \vec{p}
\right| $ \\ \hline
$R$ & $\pm 2\omega =-\bar{m}-\sqrt{\bar{m}^{2}+4\left| \vec{p}\right| ^{2}
+4\bar{m}\left( \left| \vec{p}\right| +m\right) }$ & all $\left| \vec{p}
\right| $ \\ \hline
$L$ & $\pm 2\omega =-\bar{m}+\sqrt{\bar{m}^{2}+4\left| \vec{p}\right| ^{2}
-4\bar{m}\left( \left| \vec{p}\right| -m\right) }$ & $\left| \vec{p}\right|
> $ $P_{+}$ or $\left| \vec{p}\right| < P_{-}$ \\ \hline
$L$ & $\pm 2\omega =-\bar{m}-\sqrt{\bar{m}^{2}+4\left| \vec{p}\right| ^{2}
-4\bar{m}\left( \left| \vec{p}\right| -m\right) }$ & all $\left| \vec{p}
\right| $ \\ \hline
\end{tabular}
\bigskip \newline
\noindent where $2P_{\pm }=\bar{m}\pm \sqrt{\bar{m}^{2}-4\bar{m}m}$. Here
the first column indicates whether the solution involves $\chi _{R}$ or
$\chi _{L}$, and again the upper and lower signs hold for $\psi _{R}$ and
$\psi _{L}$ respectively. The same calculation as in (4.36) of Ref. 1 shows
that the quantization condition above can be satisfied by choosing
\begin{equation}
A_{n}^{*}A_{n}=\left( \left| \omega _{n}\right| V\right) ^{-1}\left| 2\bar{m}
^{-1}\pm \omega _{n}^{-1}\right| ^{-1}
\end{equation}
in the case when there are 2 solutions for a given $\chi $ (either $\chi
_{R} $ or $\chi _{L}$) and
\begin{equation}
A_{n}^{*}A_{n}=2\left( \left| \omega _{n}\right| V\right) ^{-1}\left| 2\bar{m}
^{-1}\pm \omega _{n}^{-1}\right| ^{-1}
\end{equation}
when there is only one solution.

As in Ref. 1 (and in standard physics), when the original energy $\omega $
is negative we reinterpret the destruction operator $a$ for a particle as
the creation operator $b^{\dagger }$ for an antiparticle. If we now discard
the extremely high energy solutions, and also restrict attention to momenta
that are not extremely large, we have the following possibilities for both
sfermions and antisfermions:
\begin{center}
\begin{tabular}{|l|l|l|}
\hline
$R$ & $\omega =\left| \vec{p}\right| +m$ & all $\left| \vec{p}\right| $ \\
\hline
$L$ & $\omega =-\left| \vec{p}\right| +m$ & $\left| \vec{p}\right| <m$ \\
\hline
\end{tabular} .
\end{center}
\noindent With a Lorentz-violating mass $m$, therefore, the group velocity
$v=\partial \omega /\partial \left| \vec{p}\right| $ is $1$ for right-handed
sfermions (or antisfermions), in units with $\hbar =c=1$, and $-1$ for
left-handed sfermions. I.e., these Lorentz-violating particles have a highly
Lorentz-violating energy-momentum relation, which implies that they travel
essentially at the speed of light even though their masses are presumably
comparable to $1$ TeV. Furthermore, the velocity of a left-handed sfermion
is antiparallel to its momentum, and there are no left-handed sfermion (or
antisfermion) states for momenta with $\left| \vec{p}\right| c>mc^{2}$.

The present theory also contains gauginos etc. which can be candidates for
the dark matter, but let us suppose here that the lightest
supersymmetric partner is a neutral sfermion -- i.e., a sneutrino-- with the
highly unconventional properties discussed immediately above. For simplicity,
consider a circular orbit of radius $r$ about a mass $M$. Let us restore $c$
in the equations for clarity, and write $p=\left| \vec{p}\right| $. For
$pc\ll mc^{2}$ the general formula for the centripetal force (with 
Newtonian gravity) implies that
\begin{equation}
pv/r=GMm/r^{2} \quad \mbox{or} \quad
r=R_{S}\left( mc^{2}/2pv\right)
\end{equation}
where $R_{S}=2GM/c^{2}$ is the Schwarzschild radius. For nonrelativistic
standard cold dark matter (CDM), the kinetic energy is $pv/2$, and for
Lorentz-violating dark matter (LVDM) with $m\ll $ $\bar{m}$ it is $pv$ with
$v\approx c$.\cite{mondragon-allen-2001} For a given kinetic energy,
therefore, LVDM and CDM have orbits of comparable size in this simplistic
picture. This can also be seen from the virial theorem\cite
{mondragon-allen-2001}
$\left\langle pv\right\rangle =\left\langle \vec{p}\cdot \vec{v}\right\rangle
=-\left\langle \vec{F}\cdot \vec{r}\right\rangle =\left\langle GM\left(
r\right) m/r\right\rangle$
which implies that the kinetic energy is equal in magnitude to the
gravitational potential energy for LVDM, and to one-half the potential
energy for CDM, so dark matter with a given distribution of kinetic energies
will have comparable large-scale trajectories in both models. On the other
hand, the binding energy is vastly smaller for LVDM, and this leads to the
hope that LVDM can provide a natural explanation for the apparent
discrepancy between observations and CDM simulations in regard to cusps and
other small scale structure. There is some tentative confirmation of this
idea in preliminary computer simulations\cite{mondragon-allen-2001}.

In Lorentz-violating supergravity\cite{allen-2003}, at energies low
compared to $\bar{m}$, the fermion fields $\Psi _{f}$ and sfermion fields
$\Psi _{b}$ are coupled in the following way to the gauge fields
$A_{\mu }^{i}$, the gaugino fields $\tilde{A}_{\mu }^{i}$, the gravitational
vierbein $e_{\alpha }^{\mu }$, and the gravitino field
$\tilde{e}_{\alpha }^{\mu }$:
\begin{equation}
S_{L}=\int d^{4}x\,e\,\Psi ^{\dagger }\,iE_{\alpha }^{\mu }\sigma ^{\alpha }
\mathcal{D}_{\mu }\,\Psi \quad , \quad
\mathcal{D}_{\mu }=\partial _{\mu }-i\mathcal{A}_{\mu }^{i}t_{i} 
\end{equation}
\begin{equation}
\Psi =\left(
\begin{array}{c}
\Psi _{b} \\
\Psi _{f}
\end{array}
\right) \qquad ,\quad \mathcal{A}_{\mu }^{i}=\left(
\begin{array}{ll}
A_{\mu }^{i} & \tilde{A}_{\mu }^{i\dagger } \\
\tilde{A}_{\mu }^{i} & A_{\mu }^{i}
\end{array}
\right) \quad ,\quad E\,_{\alpha }^{\mu }=\left(
\begin{array}{ll}
e_{\alpha }^{\mu } & \tilde{e}_{\alpha }^{\mu \dagger } \\
\tilde{e}_{\alpha }^{\mu } & e_{\alpha }^{\mu }
\end{array}
\right) .
\end{equation}
We have generalized the usual vocabulary in a natural way,
so that the superpartner of the graviton is defined to be
the gravitino, and the superpartners of gauge bosons to be gauginos, even
though these fermions have quite unconventional properties. In particular,
each superpartner has both the same quantum numbers and the same spin as the
Standard Model particle. This means that the conventional R-parity is +1 for
sfermions, gauginos, and gravitinos. However, since each vertex involves
an even number of
supersymmetric partners we have conservation of an ``S-parity'', which
is -1 for sfermions, gauginos, and gravitinos, and +1 for their Standard Model
counterparts. (This is also true when the kinetic terms for the
force fields are included, as will be discussed elsewhere.) Then the lightest
supersymmetric partner (LSP) cannot spontaneously decay into lighter
Standard Model particles. For the same reason, sparticles are always created
in pairs. Parton-parton or lepton-lepton interactions will lead to
production of an even number of supersymmetric partners, just as in
standard supersymmetry with R-parity conservation. However, the sparticles
predicted by the present theory are clearly distinguished by their highly
unconventional properties.

\end{document}